\pdfoutput=1 
\documentclass{natureprintsyle}
\usepackage[utf8]{inputenc} 
\usepackage[left]{lineno} 
\usepackage{cite} 
\usepackage{graphicx} 
\usepackage{url} 
\usepackage{amsmath}	
\usepackage{amssymb}	
\usepackage[export]{adjustbox} 

\newcounter{firstbib}

\newif\iffinal
\iffinal\renewcommand{\includegraphics}[2][]{}\fi

\newcommand\aap{{Astron. Astrophys.}}
\newcommand\apjl{{Astrophys. J. Lett.}}
\newcommand\apj{{Astrophys. J.}}

\newcommand\aj{{Astron. J.}}

\newcommand\mnras{{Mon. Not. R. Astron. Soc.}}

\newcommand\pasj{{Publ. Astron. Soc. Japan}}
\newcommand\pasp{{Publ. Astron. Soc. Pacific}}

\title{A ``Heatwave'' from the G358 accretion event}
\title{Thermal heat-wave of accretion energy from the G358-MM1 high-mass protostar}

\title{A heat-wave of thermal energy from the G358-MM1 accretion burst}

\title{A heat-wave of thermal energy traced by masers in the G358-MM1 accretion burst}

\title{A heat-wave of accretion energy traced by masers in the G358-MM1 high-mass protostar}

\author{R. A. Burns,$^{1,2 \star}$, 
K. Sugiyama,$^{1,3}$
T. Hirota,$^{1}$
Kee-Tae Kim,$^{2,4}$
A. M. Sobolev,$^{5}$
B. Stecklum,$^{6}$\\
G. C. MacLeod,$^{7,8}$
Y. Yonekura,$^{9}$
M. Olech,$^{10}$
G. Orosz,$^{11,12}$
S. P. Ellingsen,$^{11}$
L. Hyland,$^{11}$\\
A. Caratti o Garatti,$^{13}$
C. Brogan,$^{14}$
T. R. Hunter,$^{14}$
C. Phillips$^{15}$
S. P. van den Heever$^{8}$\\
J. Eisl\"offel,$^{6}$
H. Linz,$^{16}$
G. Surcis,$^{17}$
J. O. Chibueze,$^{18,19}$
W. Baan,$^{20}$
B. Kramer$^{3,21}$}

\begin{document}
\maketitle

\begin{affiliations}
 \item Mizusawa VLBI Observatory, National Astronomical Observatory of Japan, 2-21-1 Osawa, Mitaka, Tokyo 181-8588, Japan
\item Korea Astronomy and Space Science Institute, 776 Daedeokdae-ro, Yuseong-gu, Daejeon 34055, Republic of Korea
\item NARIT, 260 M.4, T. Donkaew, Amphur Maerim, Chiang Mai, 50180, Thailand
\item University of Science and Technology, Korea (UST), 217 Gajeong-ro, Yuseong-gu, Daejeon 34113, Republic of Korea 
\item Ural Federal University, 19 Mira St. 620002, Ekaterinburg, Russia
\item Th\"uringer Landessternwarte, Sternwarte 5, 07778 Tautenburg, Germany
\item The University of Western Ontario, 1151 Richmond Street. London, ON N6A 3K7, Canada
\item Hartebeesthoek Radio Astronomy Observatory, PO Box 443, Krugersdorp, 1741, South Africa
\item Center for Astronomy, Ibaraki University, 2-1-1 Bunkyo, Mito, Ibaraki 310-8512, Japan
\item Centre for Astronomy, Faculty of Physics, Astronomy and Informatics, Nicolaus Copernicus University, Grudziadzka 5, 87-100 Torun, Poland
\item School of Natural Sciences, University of Tasmania, Private Bag 37, Hobart, Tasmania 7001, Australia
\item Xinjiang Astronomical Observatory, Chinese Academy of Sciences, 150 Science 1-Street, Urumqi, Xinjiang 830011, China
\item Dublin Institute for Advanced Studies, Astronomy \& Astrophysics Section, 31 Fitzwilliam Place, Dublin 2, Ireland
\item NRAO, 520 Edgemont Rd, Charlottesville, VA, 22903, USA
\item Australia Telescope National Facility, CSIRO, PO Box 76, Epping NSW 1710, Australia
\item Max Planck Institute for Astronomy, K\"onigstuhl 17, 69117 Heidelberg, Germany
\item INAF Osservatorio Astronomico di Cagliari, Via della Scienza 5, 09047 Selargius, Italy
\item Space Research Unit, Physics Department, North West University, Potchefstroom 2520, South Africa
\item Department of Physics and Astronomy, Faculty of Physical Sciences, University of Nigeria, Carver Building, 1 University Road, Nsukka, Nigeria
\item Netherlands Institute for Radio Astronomy, Dwingeloo, The Netherlands
\item Max-Planck-Institut f{\"u}r Radioastronomie, Auf dem H{\"u}gel 69, 53121 Bonn, Germany

 \item[$^{\star}$] e-mail: ross.burns@nao.ac.jp
\end{affiliations}


\begin{abstract}
High-mass stars are thought to accumulate much of their mass via short, infrequent bursts of disk-aided accretion \cite{Stamatellos11,Meyer17}. Such accretion events are rare and difficult to observe directly but are known to drive enhanced maser emission \cite{Hunter18,Gordon18,Szymczak18,Moscadelli17}. In this Letter we report high-resolution, multi-epoch methanol maser observations toward G358.93-0.03 which reveal an interesting phenomenon; the sub-luminal propagation of a thermal radiation "heat-wave" emanating from an accreting high-mass proto-star. 
The extreme transformation of the maser emission implies a sudden intensification of thermal infrared radiation from within the inner (40 mas, 270 au) region. Subsequently, methanol masers trace the radial passage of thermal radiation through the environment at $\geq$  4-8\% the speed of light. Such a high translocation rate contrasts with the $\leq$ 10 km s$^{-1}$ physical gas motions of methanol masers typically observed using very long baseline interferometry (VLBI).
The observed scenario can readily be attributed to an accretion event in the high-mass proto-star G358.93-0.03-MM1. While being the third case in its class, G358.93-0.03-MM1 exhibits unique attributes hinting at a possible `zoo' of accretion burst types.
These results promote the advantages of maser observations in understanding high-mass star formation, both through single-dish maser monitoring campaigns and via their international cooperation as VLBI arrays.
\end{abstract}


\begin{figure}[ht!]
\begin{center}
\includegraphics[width=0.369\textwidth]{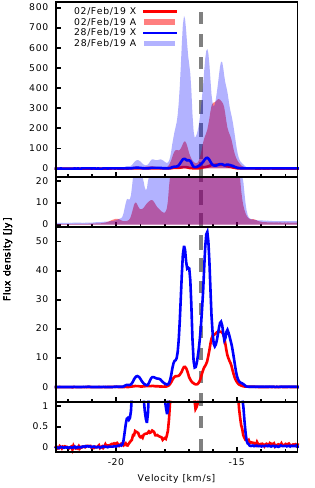}
\caption{ {\bf Spectral profiles of the 6.7 GHz methanol maser emission in G358-MM1.} Solid shapes and lines indicate the auto- and cross-correlation spectra respectively. Magnifications are shown in the lower insets to display low flux density components. The dashed line indicates the source systemic velocity. \label{SPECTRA}}
\end{center}
\end{figure}

\begin{figure*}[ht!]
\hspace{+0.27cm}
\begin{tabular}{cc}
\includegraphics[valign=T,width=0.44\textwidth]{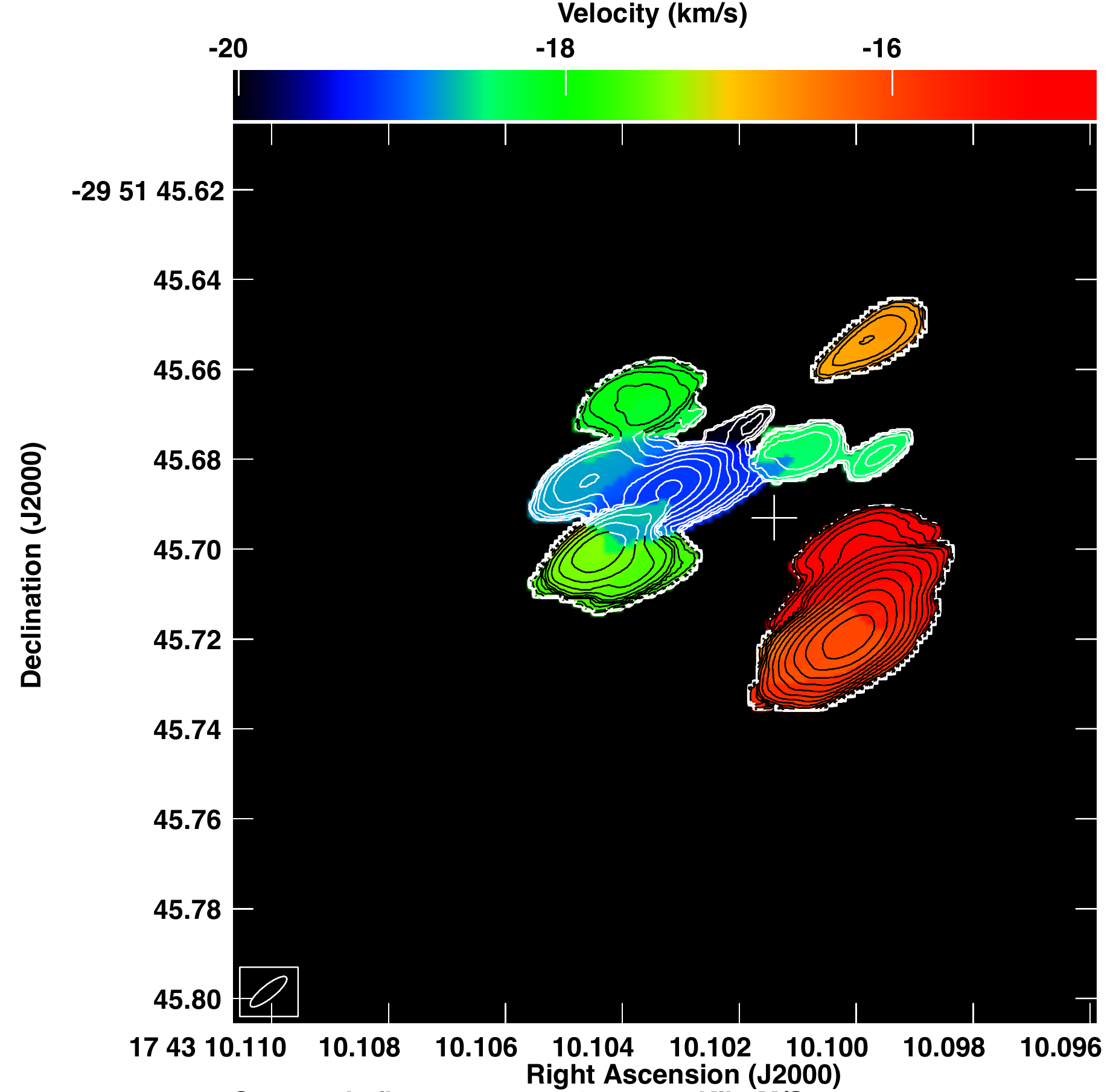} & 
\includegraphics[valign=T,width=0.44\textwidth]{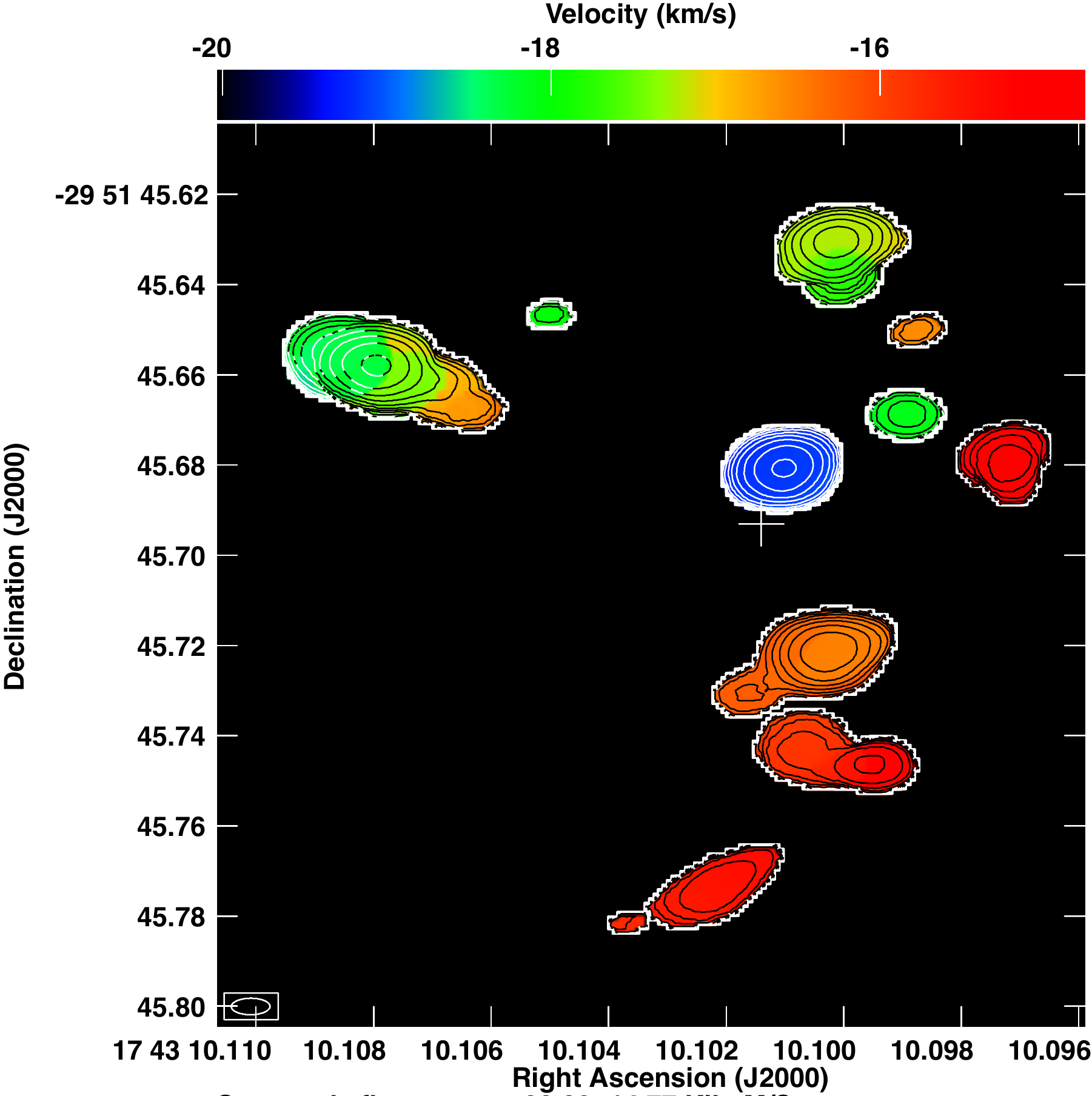}
\end{tabular}
\caption{ {\bf Methanol maser distributions in G358-MM1.} Zero'th (contours) and first (colours) moment maps of the 6.7 GHz methanol maser emission in G358-MM1. \emph{Left} shows the distribution of emission during the vx026a epoch while \emph{right} shows that of vx026c, taken 26 days later. Moment image cubes were produced for emission above a 5 $\sigma$ cutoff and contours increase by factors of 2 multiples of the first contour at 2 Jy beam$^{-1}$ km s$^{-1}$. The white cross indicates the position of the brightest millimeter continuum peak of the G358-MM1 region \cite{Brogan19a}. \label{MOMNT}}
\end{figure*}

\noindent Masers provide a novel approach to investigating accretion bursts \cite{Hunter18,Gordon18,Szymczak18,Moscadelli17}. The $5_1 \xrightarrow{} 6_0~A^+$ methanol transition at 6.7 GHz being of particular suitability as it arises in the presence of far-infrared radiation from warm ($>100$ K) dust and high gas densities (10$^{5-8}$ cm$^{-3}$) \cite{Cragg05}, making it a sign-post of high-mass star formation \cite{Breen10}. This maser has been seen to trace rotating disks and tori \cite{Bartkiewicz09,Bart16,Ellingsen06}, and its emission actively responds to changes in its local environment \cite{Fujisawa14b}.

G358.93-0.03 was discovered by its 6.7 GHz methanol maser in the Methanol Multibeam survey \cite{Caswell10a}, conducted in 2006. The early 6.7 GHz spectrum showed several $<10$ Jy peaks in the velocity range of -22.0 to -14.5 km s$^{-1}$. In January 2019 a flare of the 6.7 GHz methanol maser at -15.9 km s $^{-1}$ was identified \cite{Sugiyama19} using the Hitachi 32m telescope \cite{Yonekura16}, prompting intensive monitoring and follow-up observations across a wide range of facilities. These observations, coordinated by the Maser Monitoring Organisation (M2O, a global co-operative of maser monitoring programs. See MaserMonitoring.org) constitute the first intensive observational campaign conducted during the onset of an accretion burst in a high-mass star.

Early results from target-of-opportunity observations with the Submillimeter Array, the Atacama Large Millimeter/submillimeter Array (ALMA) \cite{Brogan19a} the Australia Telescope Compact Array (ATCA) \cite{Breen19a}, the NSF's Karl G. Jansky Very Large Array (Bayandina et al., in prep.; Chen et al., in prep.) and the Stratospheric Observatory for Infrared Astronomy (SOFIA) (Stecklum et al., in prep.) have already been established. These contemporary works have revealed striking temporal behavior \cite{MacLeod19a,Brogan19a}, rich and dynamic hot core chemistry \cite{Breen19a,Brogan19a}, complex maser emission \cite{Brogan19a}(Bayandina et al., in prep.; Chen at al., in prep.) and a kinematic signature indicating possible expansion \cite{Brogan19a}. The (sub)mm dust continuum uncovered a cluster environment of bolometric luminosity L$_{\rm bol} = 5700-22000$ L$_{\odot}$, with the most luminous source, G358.93-0.03-MM1  (hereafter "G358-MM1"), being the counterpart to the aforementioned flare activity \cite{Brogan19a}. A comparison of 160 micron flux densities measured before (HI-GAL \cite{Molinari16a}) and during the burst with FIFI-LS \cite{Fischer18a} aboard SOFIA showed an increase from $111.728 \pm 0.690$ Jy to $295.7 \pm 13.7$ Jy (Stecklum et al., in prep), roughly tripling, and verifying the occurrence of an accretion burst in G358-MM1.

Using the systemic line of sight velocity of G358-MM1 with respect to the Local Standard of Rest, $v_{\rm LSR}=-16.5 \pm 0.3$ km s$^{-1}$ \cite{Brogan19a}, gives a kinematic distance of $D=6.75^{+0.37}_{-0.68}$ kpc via the Revised kinematic Galactic distance estimation tool provided by the BeSSeL project \cite{Reid14}. Visible stars within a 0.25 arcminute field around G358 observed as part of the Gaia mission have distance estimates of $\leq 5$ kpc \cite{Bailer18}. Extinction from the G358 star forming region would impede detection of background stars, thus, the assumption that such stars are foreground to the G358 high-mass star forming region imposes a 5 kpc distance lower limit which is consistent with the kinematic distance.

\null 

As part of the M2O campaign VLBI (very long baseline interferometry) observations were initiated to provide a high angular resolution view of maser emission during the G358-MM1 maser flare. Two VLBI observations of the 6.7 GHz methanol maser were conducted with the Southern Hemmisphere long baseline array (LBA) shortly following the discovery of the maser burst (see Methods). Figure~\ref{SPECTRA} presents the auto- and cross-correlation methanol maser spectra obtained for both epochs, in which emission appears in the velocity range of -14.3 to -20.5 km s$^{-1}$; centered near the source systemic velocity \cite{Brogan19a}.

Auto-correlation spectra are sensitive to all emission within the arcminute scale primary beams of the VLBI array elements. Contrarily, cross-correlation spectra are exclusively sensitive to compact gas, on scales similar to the VLBI array synthesised beam (a few milliarcseconds). 
The similarity of the auto- and cross-correlation spectral profiles implies a common origin and similar spatio-kinematics of the maser gas at both extended and compact scales.

Comparison of the auto-correlation and cross-correlation flux densities indicate that at least 90\% of the maser emission in G358-MM1 originates from gas at angular scales larger than the synthesised beam of the LBA. Generally, maser emission in G358-MM1 rapidly evolves toward a more complex spectral profile with most features increasing in flux density by several times their initial values. 

Phase referenced, astrometric positions of the masers around G358-MM1 are shown in Figure~\ref{MOMNT}, in which a white cross indicates the peak of the millimeter core detected by ALMA at (RA, DEC) = (17:43:10.1014, -29:51:45.693; \cite{Brogan19a}). Despite notable changes across the two epochs, the 6.7 GHz methanol maser emission in G358-MM1 is generally arranged in a ring-like structure roughly centered on the position of the millimeter core. 
By the very nature of maser emission detections are biased toward regions of longest velocity-coherent path length along the line of sight. As such, the ring-like morphologies likely represent 2D projections of a 3D shell centered on G358-MM1.

\begin{figure}[ht!]
\begin{center}
\begin{tabular}{cc}
\includegraphics[valign=T,width=0.17\textwidth]{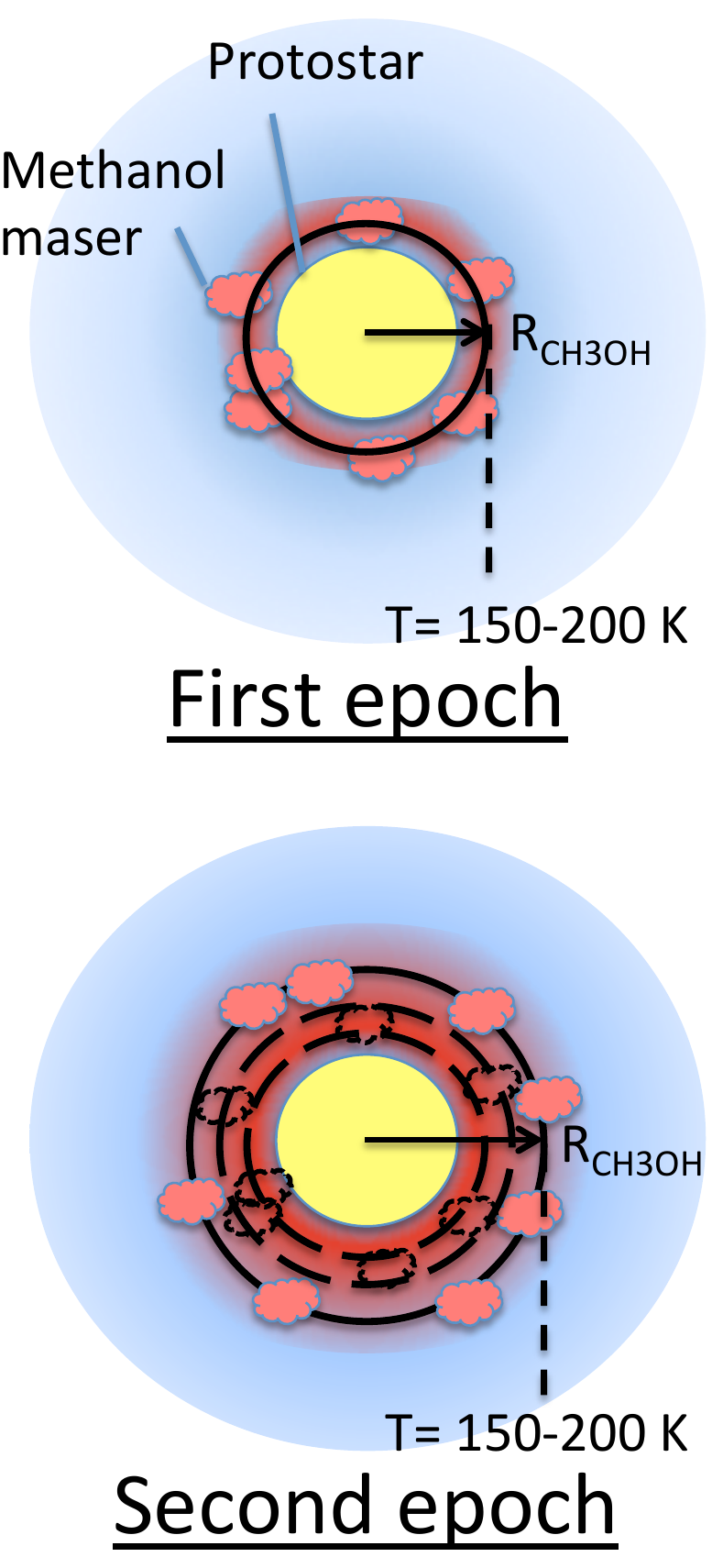} &
\hspace{-0.0cm}\includegraphics[valign=T,width=0.29\textwidth]{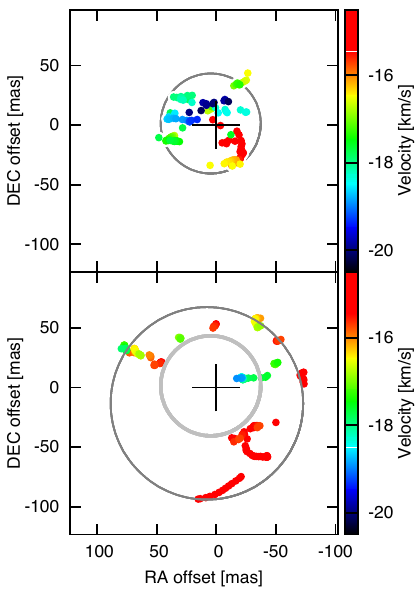}
\end{tabular}
\end{center}
\caption{ {\bf Schematic illustration of the observational data.} (\emph{Left}) A schematic model of the maser distribution and evolution in an accreting star-disk system. \emph{Right} Spot maps of emission above 5 $\sigma$ detailing the evolution of methanol maser emission in G358-MM1. Colours indicate the velocity in the frame of the local standard of rest, and symbol sizes are arbitrary. The upper and lower panels illustrate the data of vx026a (2nd Feb 2019) and vx026c (28th Feb 2019), respectively. Directional offsets are stated with respect to the coordinate (RA, DEC) = (17:43:10.1014, -29:51:45.693) which correspond to the position of G358-MM1 where the symbol size indicates the 40 mas absolute positional uncertainty in the continuum source position from \cite{Brogan19a}. The dark rings delineate the fits to each epoch, while the grey ring indicates the extent of the vx026a masers at the epoch of vx026c. \label{RINGS}}
\end{figure}

The most striking aspect of the VLBI data is the rapid transformation of the methanol maser distribution. Circles fit to the spot maps (Figure~\ref{RINGS} \emph{right}) delineate radii of 40 and 77 mas (260 and 520 au) in the first and second epochs respectively, corresponding to a translocation of 1 mas/day in the NW direction, 2 mas/day in the SE direction and an average radial expansion of 1.5 mas/day generally progressing outward from a position near the MM1 continuum peak \cite{Brogan19a}.

Considering the short period between observations (26 days), spatial evolution of the maser emission must be dominated by local phenomena ($\sim$ mas/day) with no appreciable contribution expected from Galactic systemic apparent proper motions ($\sim$ mas/yr). Translocation rates of 1-2 mas/day, correspond to $11,700$ to $23,400$ km s$^{-1}$ (0.04 to 0.08 c) at the source's kinematic distance of 6.75 kpc. We note that these sky-plane derived values are lower limits to the 3D translocation rate, which may also include a component in the line of sight direction. The gas phase abundance of methanol is relatively low in the presence of shocks faster than 10 km s$^{-1}$ \cite{Garay02a}, therefore such a fast morphological transformation can not be attributed to physical motions of methanol gas clouds. Instead, it is suitably explained by the sequential creation and quenching of maser emitting regions at ever increasing radii, as the conditions favorable to maser production \cite{Cragg05} propagate outward from an origin at G358-MM1. 

A sudden production and subsequent propagation of thermal heating, emanating from the inner 270 au region in G358-MM1, can readily be explained under the hypothesis of an accretion event in which enhanced far infrared radiation drives the production of $5_1 \xrightarrow{} 6_0~A^+$ methanol masers.
Thermal energy is transferred by photons which are scattered, absorbed and re-emitted by dust grains, consequently leading to subluminal propagation of the radiation required to produce maser emission. The mean free path of photons becomes small at high optical depths, and the resulting stagger slows down the propagation of radiative transfer. Subluminal propagation of excitation conditions was inferred for the first time by ref. \cite{Stecklum18} based on analyses of the S255IR-NIRS3 maser burst \cite{Moscadelli17}.

Alternative explanations for the phenomena seen in G358-MM1, including super-/nova and changes in the geometry and/or radiation of a high-mass protostellar system fail to simultaneously explain the global flux increase in all velocity components of the maser emission, its temporal behavior and rapid morphological changes, in addition to the accompanying FIR enhancement. Contrarily, all of which are consistent with the accretion burst hypothesis.

High-mass protostars may be facilitated in achieving their necessarily high accretion rates by disk-aided inward transport of material and a reduction of stellar UV radiation. The latter may be induced via non-continuous accretion mechanisms \cite{Audard14} which cause a bloating of the stellar radius and a subsequent reduction in its effective temperature \cite{Hosokawa10}.
Such `episodic accretion' is exemplified by the FUori and EXori class of low-mass protostars (reviewed in \cite{Audard14}) and, despite differences in the initial conditions, timescales, environments and protostellar radiation fields present in low- and high-mass protostars \cite{Tan16a}, episodic accretion is quickly becoming considered a necessary component of high-mass star formation \cite{Stamatellos11,Meyer17}. Accretion bursts were identified in high-mass protostars S255IR-NIRS3 \cite{Garatti17} and NGC6334I-MM1 \cite{Hunter17} and the behavior has been inferred indirectly via outflow ejection histories \cite{Burns16b}.

The G358-MM1 accretion burst contrasts to those of S255IR-NIRS3 \cite{Garatti17} and NGC6334I-MM1 \cite{Hunter17} in that no clear enhancements in the millimetric continuum and near-infrared were confirmed \cite{Brogan19a,Stecklum19a}. The former can be explained by a high visual extinction in the immediate surroundings of the high-mass protostar, and the latter implying a subsequent failure to sufficiently heat the dust in the disk mid-plane. This may be due to a less substantial mass transfer, leading to a less intense accretion luminosity.
Indeed, S255IR-NIRS3 exhibited a bolometric luminosity increase of a factor of 4 while that of NGC6334I increased by a factor of 10, despite a common underlying mechanism. Furthermore, the flux of the G358 6.7 GHz methanol maser was much fainter ($\sim 8$ Jy) in its quiescent phase compared to those of S255IR-NIRS3 (46.5 Jy \cite{Szymczak18}) and NGC6334I (>2000 Jy \cite{Gordon18}), and the maser burst exhibited a quicker rise and decline in the case of G358-MM1.

The G358-MM1 event may therefore represent a new species amongst a `zoo' of high-mass protostar accretion burst varieties. It is likely that this class of events will diversify as more follow-up investigations of accretion bursts are reported. Maser monitoring observations will continue to uncover new accretion burst candidates in high-mass protostars and VLBI will provide the investigative keys to understanding the physical processes taking place.






\newpage

\section*{Methods}
~
Target of opportunity observations were requested to the LBA on January the 28th 2019 in response to reports from the M2O of the 6.7 GHz burst event \cite{Sugiyama19_2}. Two VLBI epochs were granted, the first was conducted on the 2nd of Feb 2019 with a VLBI array consisting of stations: Ceduna, Hobart, Mopra, Warkworth, Hartebeesthoek and the ATCA in phased array mode. The second epoch on the 28th of Feb 2019 consisted of Ceduna, Hobart, Mopra, Warkworth, the ATCA and Parkes.  Observations were conducted under project codes vx026a and vx026c for the first and second epochs, respectively.

Observations consisted of phase reference cycles between G358-MM1 and two phase reference sources, J1744-3116 and J1743-3058, interleaved, with a 4.5 min cycle time. Each hour, scans were made of one of the bright calibration sources, 3C273, PKS B1934-638, NRAO530. Two 16 MHz channels of dual circular polarisation data were recorded with 2 bit, Nyquist sampling, corresponding to a data rate of 128 Mbps. The total observing times in vx026a and vx026c were 11 and 13 hours respectively.

Data were correlated with a 2 second accumulation period at the Pawsey correlator center using DiFX \cite{Deller07}. Two passes were performed, one comprising the full bandwidth correlated at 32 spectral points per channel, and one single `zoom band' of 4 MHz, centered on the maser peak, with 4096 spectral points corresponding to frequency and velocity spacings of 0.977 kHz, and 0.045 km s$^{-1}$. Full Stokes correlation products were computed.

Data were calibrated using the AIPS software package. Individual station gains were calibrated by scaling the auto-correlation spectrum of the maser to match contemporaneous monitoring data provided by the M2O, using the AIPS task ACFIT. Absolute flux density calibration is considered accurate to 20\%. Bandpass subtracted auto- and cross-correlated spectra for both epochs are shown in Figure~\ref{SPECTRA}.

Delay calibration was performed using 3C273, PKS B1934-638 and NRAO530 and solutions applied to all sources. Phase calibration was established using the maser data and refined by self calibration to remove the effects of source structure. The solutions were then applied to J1744-3116 and J1743-3058, thereby establishing the coordinates of the maser source with respect to the International Celestial Reference Frame (ICRF). Comparison of the astrometric positions of the two ICRF quasars allowed evaluation of the systematic positional offsets by fixing one (J1744-3116) and using the other (J1743-3058) as a check source. Astrometric accuracy was reliable to within $\pm 3$ milli-arcseconds (mas). During imaging the data were weighted to produce a comparable synthesised beam shape across the two epochs, these were $10.17 \times 2.96$ and $8.56 \times 3.64$ mas for vx026a and vx026c, respectively. Final image noise rms values were elevated in those containing bright emission where the maximum noise measured in the brightest channels were 80 and 210 mJy for vx026a and vx026c, respectively.

Zero'th and first moment image cubes were produced for both epochs in which a channel dependent noise cutoff of 5 $\sigma$ was applied to remove strong side-lobe emission. These are presented as coloured contour maps in Figure~\ref{MOMNT} in which contours increase in factors of 2 from the first contour at 2 Jy beam$^{-1}$ km s$^{-1}$.
Finally, two dimensional Gaussian functions were fit to maser emission above a noise cutoff of 5 $\sigma$ in all frequency channels to produce the spot maps shown in Figure~\ref{RINGS} where symbol colours represent the LSR velocity and spot sizes are arbitrary. The slight differences in the maser distributions seen in Figures~\ref{MOMNT} and \ref{RINGS} comes from the omission of regions below the 2 Jy beam$^{-1}$ km s$^{-1}$ first contour in the former.

\null 

\subsection{Data availability.}

The data that support the plots within this paper and other findings of this study are available from the PAWSEY data archive\\
(https://data.pawsey.org.au/public/?path=/VLBI/Archive/LBA/vx026) or from the corresponding author upon reasonable request



\begin{addendum}
\item[Acknowledgements] 

R.A.B. acknowledges support through the EACOA Fellowship from the East Asian Core Observatories Association. S.P.E., G.O. and L.H. acknowledge the support of the ARC Discovery Project (project number DP180101061). G.O. was supported by CAS LCWR grant 2018-XBQNXZ-B-021. A.M.S. was supported by the Foundation for the Advancement of Theoretical Physics and Mathematics “BASIS". A.M.S. is supported by the Ministry of Science and High Education (the basic part of the state assignment, RK No. AAAA-A17-117030310283-7) and by the Act 211 Government of the Russian Federation, contract 02.A03.21.0006. This work was supported by JSPS KAKENHI grant JP19K03921. T.H. is financially supported by the MEXT/JSPS KAKENHI grants 16K05293 and 17K05398. J.O.C. acknowledges support by the Italian Ministry of Foreign Affairs and International Cooperation (MAECI Grant Number ZA18GR02) and the South African Department of Science and Technology’s National Research Foundation (DST-NRF Grant Number 113121) as part of the ISARP RADIOSKY2020 Joint Research Scheme. This work was supported by the National Science Centre, Poland, through grant 2016/21/B/ST9/01455. The work was also supported by resources provided by the Pawsey Supercomputing Centre with funding from the Australian Government and the Government of Western Australia. The LBA is part of the Australia Telescope National Facility, which is funded by the Australian Government for operation as a National Facility managed by CSIRO. This work was supported by resources provided by the Pawsey Supercomputing Centre with funding from the Australian Government and the Government of Western Australia. The National Radio Astronomy Observatory is a facility of the National Science Foundation operated under cooperative agreement by Associated Universities, Inc.

\newpage 

\item[Author contributions] R.A.B. lead the project as principle investigator for the observations, processed the data, and authored the manuscript. K.S. and Y.Y. selected the target maser source. B.S., J.E., A.C.G and A.M.S provided theoretical interpretations of the data. G.O, S.P.E, L.H and C.P. conducted the observations. All authors assisted in the interpretation of the results and contributed to the preparation of the manuscript.

\item[Additional information]
{\bf~\\Supplementary information} is available for this paper.
{\bf~\\Reprints and permissions information} is available at \url{www.nature.com/reprints}.
{\bf~\\Correspondence and requests for materials} should be address to R.A.Burns (ross.burns@nao.ac.jp).

\item[Competing interests] The authors declare no competing financial interests.

\noindent This manuscript was prepared using the author's custom LaTeX template. Research contents reflect the state of the paper before proofs. The final article published by Springer Nature should be taken as the most complete version.

\end{addendum}


\section*{References}

\begin{thebibliography}{10}
\expandafter\ifx\csname url\endcsname\relax
  \def\url#1{\texttt{#1}}\fi
\expandafter\ifx\csname urlprefix\endcsname\relax\def\urlprefix{URL }\fi
\providecommand{\bibinfo}[2]{#2}
\providecommand{\eprint}[2][]{\url{#2}}

\bibitem{Stamatellos11}
\bibinfo{author}{{Stamatellos}, D.}, \bibinfo{author}{{Whitworth}, A.~P.} \&
  \bibinfo{author}{{Hubber}, D.~A.}
\newblock \bibinfo{title}{{The Importance of Episodic Accretion for Low-mass
  Star Formation}}.
\newblock \emph{\bibinfo{journal}{\apj}} \textbf{\bibinfo{volume}{730}},
  \bibinfo{pages}{32} (\bibinfo{year}{2011}).
\newblock \eprint{1103.1378}.

\bibitem{Meyer17}
\bibinfo{author}{{Meyer}, D.~M.-A.}, \bibinfo{author}{{Vorobyov}, E.~I.},
  \bibinfo{author}{{Kuiper}, R.} \& \bibinfo{author}{{Kley}, W.}
\newblock \bibinfo{title}{{On the existence of accretion-driven bursts in
  massive star formation}}.
\newblock \emph{\bibinfo{journal}{\mnras}} \textbf{\bibinfo{volume}{464}},
  \bibinfo{pages}{L90--L94} (\bibinfo{year}{2017}).
\newblock \eprint{1609.03402}.

\bibitem{Hunter18}
\bibinfo{author}{{Hunter}, T.~R.} \emph{et~al.}
\newblock \bibinfo{title}{{The Extraordinary Outburst in the Massive
  Protostellar System NGC 6334I-MM1: Emergence of Strong 6.7 GHz Methanol
  Masers}}.
\newblock \emph{\bibinfo{journal}{\apj}} \textbf{\bibinfo{volume}{854}},
  \bibinfo{pages}{170} (\bibinfo{year}{2018}).
\newblock \eprint{1801.02141}.

\bibitem{Gordon18}
\bibinfo{author}{{MacLeod}, G.~C.} \emph{et~al.}
\newblock \bibinfo{title}{{A masing event in NGC 6334I: contemporaneous flaring
  of hydroxyl, methanol, and water masers}}.
\newblock \emph{\bibinfo{journal}{\mnras}} \textbf{\bibinfo{volume}{478}},
  \bibinfo{pages}{1077--1092} (\bibinfo{year}{2018}).
\newblock \eprint{1804.05308}.

\bibitem{Szymczak18}
\bibinfo{author}{{Szymczak}, M.}, \bibinfo{author}{{Olech}, M.},
  \bibinfo{author}{{Wolak}, P.}, \bibinfo{author}{{G{\'e}rard}, E.} \&
  \bibinfo{author}{{Bartkiewicz}, A.}
\newblock \bibinfo{title}{{Giant burst of methanol maser in S255IR-NIRS3}}.
\newblock \emph{\bibinfo{journal}{\aap}} \textbf{\bibinfo{volume}{617}},
  \bibinfo{pages}{A80} (\bibinfo{year}{2018}).
\newblock \eprint{1807.07334}.

\bibitem{Moscadelli17}
\bibinfo{author}{{Moscadelli}, L.} \emph{et~al.}
\newblock \bibinfo{title}{{Extended CH$_{3}$OH maser flare excited by a
  bursting massive YSO}}.
\newblock \emph{\bibinfo{journal}{\aap}} \textbf{\bibinfo{volume}{600}},
  \bibinfo{pages}{L8} (\bibinfo{year}{2017}).

\bibitem{Cragg05}
\bibinfo{author}{{Cragg}, D.~M.}, \bibinfo{author}{{Sobolev}, A.~M.} \&
  \bibinfo{author}{{Godfrey}, P.~D.}
\newblock \bibinfo{title}{{Models of class II methanol masers based on improved
  molecular data}}.
\newblock \emph{\bibinfo{journal}{\mnras}} \textbf{\bibinfo{volume}{360}},
  \bibinfo{pages}{533--545} (\bibinfo{year}{2005}).
\newblock \eprint{astro-ph/0504194}.

\bibitem{Breen10}
\bibinfo{author}{{Breen}, S.~L.}, \bibinfo{author}{{Ellingsen}, S.~P.},
  \bibinfo{author}{{Caswell}, J.~L.} \& \bibinfo{author}{{Lewis}, B.~E.}
\newblock \bibinfo{title}{{12.2-GHz methanol masers towards 1.2-mm dust clumps:
  quantifying high-mass star formation evolutionary schemes}}.
\newblock \emph{\bibinfo{journal}{\mnras}} \textbf{\bibinfo{volume}{401}},
  \bibinfo{pages}{2219--2244} (\bibinfo{year}{2010}).
\newblock \eprint{0910.1223}.

\bibitem{Bartkiewicz09}
\bibinfo{author}{{Bartkiewicz}, A.}, \bibinfo{author}{{Szymczak}, M.},
  \bibinfo{author}{{van Langevelde}, H.~J.}, \bibinfo{author}{{Richards},
  A.~M.~S.} \& \bibinfo{author}{{Pihlstr{\"o}m}, Y.~M.}
\newblock \bibinfo{title}{{The diversity of methanol maser morphologies from
  VLBI observations}}.
\newblock \emph{\bibinfo{journal}{\aap}} \textbf{\bibinfo{volume}{502}},
  \bibinfo{pages}{155--173} (\bibinfo{year}{2009}).
\newblock \eprint{0905.3469}.

\bibitem{Bart16}
\bibinfo{author}{{Bartkiewicz}, A.}, \bibinfo{author}{{Szymczak}, M.} \&
  \bibinfo{author}{{van Langevelde}, H.~J.}
\newblock \bibinfo{title}{{European VLBI Network imaging of 6.7 GHz methanol
  masers}}.
\newblock \emph{\bibinfo{journal}{\aap}} \textbf{\bibinfo{volume}{587}},
  \bibinfo{pages}{A104} (\bibinfo{year}{2016}).
\newblock \eprint{1601.03197}.

\bibitem{Ellingsen06}
\bibinfo{author}{{Ellingsen}, S.~P.}
\newblock \bibinfo{title}{{Methanol Masers: Reliable Tracers of the Early
  Stages of High-Mass Star Formation}}.
\newblock \emph{\bibinfo{journal}{\apj}} \textbf{\bibinfo{volume}{638}},
  \bibinfo{pages}{241--261} (\bibinfo{year}{2006}).
\newblock \eprint{astro-ph/0510218}.

\bibitem{Fujisawa14b}
\bibinfo{author}{{Fujisawa}, K.} \emph{et~al.}
\newblock \bibinfo{title}{{Observations of the bursting activity of the 6.7 GHz
  methanol maser in G33.641-0.228}}.
\newblock \emph{\bibinfo{journal}{\pasj}} \textbf{\bibinfo{volume}{66}},
  \bibinfo{pages}{109} (\bibinfo{year}{2014}).
\newblock \eprint{1408.3695}.

\bibitem{Caswell10a}
\bibinfo{author}{{Caswell}, J.~L.} \emph{et~al.}
\newblock \bibinfo{title}{{The 6-GHz methanol multibeam maser catalogue - I.
  Galactic Centre region, longitudes 345$^{\circ}$ to 6$^{\circ}$}}.
\newblock \emph{\bibinfo{journal}{\mnras}} \textbf{\bibinfo{volume}{404}},
  \bibinfo{pages}{1029--1060} (\bibinfo{year}{2010}).
\newblock \eprint{1002.2475}.

\bibitem{Sugiyama19}
\bibinfo{author}{{Sugiyama}, K.}, \bibinfo{author}{{Saito}, Y.},
  \bibinfo{author}{{Yonekura}, Y.} \& \bibinfo{author}{{Momose}, M.}
\newblock \bibinfo{title}{{Bursting activity of the 6.668-GHz CH3OH maser
  detected in G 358.93-00.03 using the Hitachi 32-m}}.
\newblock \emph{\bibinfo{journal}{ATEL}} \textbf{\bibinfo{volume}{12446}}
  (\bibinfo{year}{2019}).

\bibitem{Yonekura16}
\bibinfo{author}{{Yonekura}, Y.} \emph{et~al.}
\newblock \bibinfo{title}{{The Hitachi and Takahagi 32 m radio telescopes:
  Upgrade of the antennas from satellite communication to radio astronomy}}.
\newblock \emph{\bibinfo{journal}{\pasj}} \textbf{\bibinfo{volume}{68}},
  \bibinfo{pages}{74} (\bibinfo{year}{2016}).

\bibitem{Brogan19a}
\bibinfo{author}{{Brogan}, C.~L.} \emph{et~al.}
\newblock \bibinfo{title}{{Sub-arcsecond (Sub)millimeter Imaging of the Massive
  Protocluster G358.93{\ensuremath{-}}0.03: Discovery of 14 New Methanol Maser
  Lines Associated with a Hot Core}}.
\newblock \emph{\bibinfo{journal}{\apjl}} \textbf{\bibinfo{volume}{881}},
  \bibinfo{pages}{L39} (\bibinfo{year}{2019}).
\newblock \eprint{1907.02470}.

\bibitem{Breen19a}
\bibinfo{author}{{Breen}, S.~L.} \emph{et~al.}
\newblock \bibinfo{title}{{Discovery of Six New Class II Methanol Maser
  Transitions, Including the Unambiguous Detection of Three Torsionally Excited
  Lines toward G 358.931-0.030}}.
\newblock \emph{\bibinfo{journal}{\apj}} \textbf{\bibinfo{volume}{876}},
  \bibinfo{pages}{L25} (\bibinfo{year}{2019}).
\newblock \eprint{1904.06853}.


\bibitem{Stecklum19a}
\bibinfo{author}{{Stecklum}, B.} \emph{et~al.}
\newblock \bibinfo{title}{{in prep.}}
\newblock \emph{\bibinfo{journal}{,}}  (\bibinfo{year}{2019}).

\bibitem{MacLeod19a}
\bibinfo{author}{{MacLeod}, G.~C.} \emph{et~al.}
\newblock \bibinfo{title}{{Submitted to}}.
\newblock \emph{\bibinfo{journal}{\mnras}}  (\bibinfo{year}{2019}).

\bibitem{Molinari16a}
\bibinfo{author}{{Molinari}, S.} \emph{et~al.}
\newblock \bibinfo{title}{{Hi-GAL, the Herschel infrared Galactic Plane Survey:
  photometric maps and compact source catalogues. First data release for the
  inner Milky Way: +68$^{\circ}$ {\ensuremath{\geq}} l {\ensuremath{\geq}}
  -70$^{\circ}$}}.
\newblock \emph{\bibinfo{journal}{\aap}} \textbf{\bibinfo{volume}{591}},
  \bibinfo{pages}{A149} (\bibinfo{year}{2016}).
\newblock \eprint{1604.05911}.

\bibitem{Fischer18a}
\bibinfo{author}{{Fischer}, C.} \emph{et~al.}
\newblock \bibinfo{title}{{FIFI-LS: The Field-Imaging Far-Infrared Line
  Spectrometer on SOFIA}}.
\newblock \emph{\bibinfo{journal}{Journal of Astronomical Instrumentation}}
  \textbf{\bibinfo{volume}{7}}, \bibinfo{pages}{1840003--556}
  (\bibinfo{year}{2018}).

\bibitem{Reid14}
\bibinfo{author}{{Reid}, M.~J.} \emph{et~al.}
\newblock \bibinfo{title}{{Trigonometric Parallaxes of High Mass Star Forming
  Regions: The Structure and Kinematics of the Milky Way}}.
\newblock \emph{\bibinfo{journal}{\apj}} \textbf{\bibinfo{volume}{783}},
  \bibinfo{pages}{130} (\bibinfo{year}{2014}).
\newblock \eprint{1401.5377}.


\bibitem{Bailer18}
\bibinfo{author}{{Bailer-Jones}, C.~A.~L.}, \bibinfo{author}{{Rybizki}, J.},
  \bibinfo{author}{{Fouesneau}, M.}, \bibinfo{author}{{Mantelet}, G.} \&
  \bibinfo{author}{{Andrae}, R.}
\newblock \bibinfo{title}{{Estimating Distance from Parallaxes. IV. Distances
  to 1.33 Billion Stars in Gaia Data Release 2}}.
\newblock \emph{\bibinfo{journal}{\aj}} \textbf{\bibinfo{volume}{156}},
  \bibinfo{pages}{58} (\bibinfo{year}{2018}).
\newblock \eprint{1804.10121}.

\bibitem{Garay02a}
\bibinfo{author}{{Garay}, G.}, \bibinfo{author}{{Mardones}, D.},
  \bibinfo{author}{{Rodr{\'{\i}}guez}, L.~F.}, \bibinfo{author}{{Caselli}, P.}
  \& \bibinfo{author}{{Bourke}, T.~L.}
\newblock \bibinfo{title}{{Methanol and Silicon Monoxide Observations toward
  Bipolar Outflows Associated with Class 0 Objects}}.
\newblock \emph{\bibinfo{journal}{\apj}} \textbf{\bibinfo{volume}{567}},
  \bibinfo{pages}{980--998} (\bibinfo{year}{2002}).

\bibitem{Stecklum18}
\bibinfo{author}{{Stecklum}, B.} \emph{et~al.}
\newblock \bibinfo{title}{{Infrared variability, maser activity, and accretion
  of massive young stellar objects}}.
\newblock In \bibinfo{editor}{{Tarchi}, A.}, \bibinfo{editor}{{Reid}, M.~J.} \&
  \bibinfo{editor}{{Castangia}, P.} (eds.)
  \emph{\bibinfo{booktitle}{Astrophysical Masers: Unlocking the Mysteries of
  the Universe}}, vol. \bibinfo{volume}{336} of \emph{\bibinfo{series}{IAU
  Symposium}}, \bibinfo{pages}{37--40} (\bibinfo{year}{2018}).
\newblock \eprint{1711.01489}.

\bibitem{Audard14}
\bibinfo{author}{{Audard}, M.} \emph{et~al.}
\newblock \bibinfo{title}{{Episodic Accretion in Young Stars}}.
\newblock \emph{\bibinfo{journal}{Protostars and Planets VI}}
  \bibinfo{pages}{387--410} (\bibinfo{year}{2014}).
\newblock \eprint{1401.3368}.

\bibitem{Hosokawa10}
\bibinfo{author}{{Hosokawa}, T.}, \bibinfo{author}{{Yorke}, H.~W.} \&
  \bibinfo{author}{{Omukai}, K.}
\newblock \bibinfo{title}{{Evolution of Massive Protostars Via Disk
  Accretion}}.
\newblock \emph{\bibinfo{journal}{\apj}} \textbf{\bibinfo{volume}{721}},
  \bibinfo{pages}{478--492} (\bibinfo{year}{2010}).
\newblock \eprint{1005.2827}.

\bibitem{Tan16a}
\bibinfo{author}{{Tan}, J.~C.}
\newblock \bibinfo{title}{{Comparison of Low-Mass and High-Mass Star
  Formation}}.
\newblock In \bibinfo{editor}{{Jablonka}, P.}, \bibinfo{editor}{{Andr{\'e}},
  P.} \& \bibinfo{editor}{{van der Tak}, F.} (eds.)
  \emph{\bibinfo{booktitle}{From Interstellar Clouds to Star-Forming Galaxies:
  Universal Processes?}}, vol. \bibinfo{volume}{315} of
  \emph{\bibinfo{series}{IAU Symposium}}, \bibinfo{pages}{154--162}
  (\bibinfo{year}{2016}).
\newblock \eprint{1510.08021}.

\bibitem{Garatti17}
\bibinfo{author}{{Caratti o Garatti}, A.} \emph{et~al.}
\newblock \bibinfo{title}{{Disk-mediated accretion burst in a high-mass young
  stellar object}}.
\newblock \emph{\bibinfo{journal}{Nature Physics}}
  \textbf{\bibinfo{volume}{13}}, \bibinfo{pages}{276--279}
  (\bibinfo{year}{2017}).
\newblock \eprint{1704.02628}.

\bibitem{Hunter17}
\bibinfo{author}{{Hunter}, T.~R.} \emph{et~al.}
\newblock \bibinfo{title}{{An Extraordinary Outburst in the Massive
  Protostellar System NGC6334I-MM1: Quadrupling of the Millimeter Continuum}}.
\newblock \emph{\bibinfo{journal}{\apjl}} \textbf{\bibinfo{volume}{837}},
  \bibinfo{pages}{L29} (\bibinfo{year}{2017}).
\newblock \eprint{1701.08637}.

\bibitem{Burns16b}
\bibinfo{author}{{Burns}, R.~A.}, \bibinfo{author}{{Handa}, T.},
  \bibinfo{author}{{Nagayama}, T.}, \bibinfo{author}{{Sunada}, K.} \&
  \bibinfo{author}{{Omodaka}, T.}
\newblock \bibinfo{title}{{H2O masers in a jet-driven bow shock: episodic
  ejection from a massive young stellar object}}.
\newblock \emph{\bibinfo{journal}{\mnras}} \textbf{\bibinfo{volume}{460}},
  \bibinfo{pages}{283--290} (\bibinfo{year}{2016}).


\setcounter{firstbib}{\value{enumiv}}

\end{thebibliography}

\section*{References}

\begin{thebibliography}{10}
\expandafter\ifx\csname url\endcsname\relax
  \def\url#1{\texttt{#1}}\fi
\expandafter\ifx\csname urlprefix\endcsname\relax\def\urlprefix{URL }\fi
\providecommand{\bibinfo}[2]{#2}
\providecommand{\eprint}[2][]{\url{#2}}

\addtocounter{enumiv}{\value{firstbib}}

\bibitem{Deller07}
\bibinfo{author}{{Deller}, A.~T.}, \bibinfo{author}{{Tingay}, S.~J.},
  \bibinfo{author}{{Bailes}, M.} \& \bibinfo{author}{{West}, C.}
\newblock \bibinfo{title}{{DiFX: A Software Correlator for Very Long Baseline
  Interferometry Using Multiprocessor Computing Environments}}.
\newblock \emph{\bibinfo{journal}{\pasp}} \textbf{\bibinfo{volume}{119}},
  \bibinfo{pages}{318--336} (\bibinfo{year}{2007}).
\newblock \eprint{astro-ph/0702141}.

\bibitem{Sugiyama19_2}
\bibinfo{author}{{Sugiyama}, K.}, \bibinfo{author}{{Saito}, Y.},
  \bibinfo{author}{{Yonekura}, Y.} \& \bibinfo{author}{{Momose}, M.}
\newblock \bibinfo{title}{{Bursting activity of the 6.668-GHz CH3OH maser
  detected in G 358.93-00.03 using the Hitachi 32-m}}.
\newblock \emph{\bibinfo{journal}{ATEL}} \textbf{\bibinfo{volume}{12446}}
  (\bibinfo{year}{2019}).


\end{thebibliography}
\end{document}